\documentclass[english,aps,pre,showpacs,preprint]{revtex4}
\usepackage{amsmath}
\usepackage{amssymb}
\usepackage{graphicx}
\usepackage{epsfig}
\usepackage{babel}

\makeatletter

\input{tcilatex}
\makeatother

\begin{document}

\title{Field Theory of Polymers: Escaping the Sign Problem}
\author{Glenn H. Fredrickson}
\email{ghf@mrl.ucsb.edu}
\affiliation{Departments of Chemical Engineering \& Materials and Materials Research
Laboratory, University of California, Santa Barbara, CA 93106, USA}
\author{Henri Orland}
\email{orland@spht.saclay.cea.fr}
\affiliation{Service de Physique Theorique, CE-Saclay, CEA 91191 Gif-sur-Yvette Cedex,
France}
\date{\today{}}

\begin{abstract}
We examine statistical field theories of polymeric fluids in view of
performing numerical simulations. The partition function of these systems
can be expressed as a functional integral over real density fields. The
introduction of density field variables serves to decouple interactions
among non-bonded monomers, and renders the resulting effective Hamiltonian $%
H $ for the field theory real and the Boltzmann factor $\exp (-H)$
positive definite. This is in contrast with conventional (Edwards
type) field theories expressed in terms of chemical potentials
that have complex $H$. The density field theory involves the
calculation of an intermediate functional integral, which provides
the entropy of the polymer fluid at a fixed density profile. This
functional integral is positive definite and we show that in the
thermodynamic limit of large systems, it is dominated by saddle
points of the integrand. This procedure side-steps the
\textquotedblleft sign problem\textquotedblright\ in the chemical
potential field formulation. The formalism is illustrated in the
context of models of flexible polymers. We discuss the
implications for field-theoretic computer simulations of polymeric
fluids.
\end{abstract}

\pacs{}
\maketitle

\newpage

\section{Introduction}

Field theories have proved to be a useful theoretical framework for studying
the equilibrium and time-dependent properties of a wide variety of complex
fluid systems. In the 1960s, Edwards introduced field-theoretic methods to
the field of polymer physics and applied these techniques fruitfully to
examine the universal aspects of chain conformational statistics and
thermodynamic properties of polymer solutions \cite{edwards65}. Perturbation
methods, mean-field approximations, and variational schemes were introduced
for tackling important problems such as the excluded volume effect and
addressing collective phenomena, such as screening in semi-dilute solutions.
These analytical methods were considerably enhanced in the 1970s and 1980s
with the advent of renormalization group theory, the introduction of scaling
concepts, and the discovery of the connection of polymer statistical
mechanics with the $n$-vector field theory model for $n\rightarrow 0$ \cite%
{degennes}.

Unfortunately, most of these analytical techniques are difficult
to apply in the important case of \emph{inhomogeneous} polymers,
which include multicomponent polymer alloys, block and graft
copolymers, nano-composites, thin films, emulsions, and
suspensions. Appropriate field theory models for such systems are
straightforward to construct, but the analytical tools available
for examining the properties of the models are rather limited. One
powerful tool is self-consistent field theory (SCFT), which
amounts to a saddle point approximation to the partition function
of the field theory \cite{helfand75,schmid98}. However, the SCFT
field equations determining the ``mean fields'' are nonlinear,
nonlocal, and defy analytical solution. Certain limiting cases of
SCFT, such as weak inhomogeneities or fields that vary slowly over
the size of a polymer, and for simple geometries, are analytically
tractable. In such special circumstances, however, the SCFT
solutions still reflect the mean-field assumption intrinsic to the
theory, which further restricts their applicability. In
particular, such solutions are primarily limited to the
description of concentrated solutions or melts of high molecular
weight polymers. Even in the case of dense molten blends or
copolymers, SCFT can fail in situations where soft modes appear,
such as near critical points, unbinding transitions, or in
micellar or microemulsion phases \cite{bates97}, since
thermally-induced field fluctuations can then be significant.

In recent years, it has become apparent that field theory models of polymers
and complex fluids, besides serving as the platform for analytical
calculations, can also provide a flexible framework for numerical
simulations \cite{ghfrev}. Building on significant advances in numerical
methods for solving the mean-field SCFT equations \cite{fraaije97,matsen94},
techniques have been developed for direct numerical sampling of field
fluctuations \cite{ganesan01,ghfrev}. These methods open many exciting
possibilities for carrying out ``field-theoretic simulations'' of a wide
variety of inhomogeneous polymer systems. Aside from the numerical errors
associated with representing the fields and the usual statistical errors
intrinsic to any computer simulation technique, field-theoretic simulations
provide a route to studying the exact properties of a polymer fluid model in
the absence of any simplifying approximations. Such field-based simulations
also facilitate systematic coarse-graining, making it possible to
equilibrate much larger systems with a broader range of structural scales
than in conventional atomistic computer simulations. At the present time the
methods are limited to \emph{equilibrium } structure and properties,
although research is underway to extend them to nonequilibrium situations.

A difficulty with the current strategies for field-based simulations of
polymeric fluids is that they rely on the introduction of
Hubbard-Stratonovich ``chemical potential'' fields, first employed for
polymers by Edwards \cite{edwards65}, to decouple the non-bonded
interactions among monomers. This allows one to explicitly trace out the
monomer coordinates in favor of the introduced fields, generating a field
theory model for a polymer solution or melt. Unfortunately, this description
leads to an effective Hamiltonian $\mathcal{G} (\phi )$, expressing the
energy of a chemical potential field configuration $\phi (r)$, that is \emph{%
complex}, rather than real. As a consequence, the Boltzmann factor $\exp (-%
\mathcal{G})$ is not positive definite, so conventional Monte Carlo sampling
of the partition function is problematic \cite{lin86}. This difficulty in
sampling field configurations is a special case of a more general ``sign
problem'', which involves the development of techniques for efficient
numerical evaluation of high dimensional integrals with non-positive
definite integrands. While promising new methods have appeared for battling
the sign problem \cite{baeurle02,ghf03}, the problem is by no means
``solved''. Oscillations in the phase factor $\exp (- i \: \mathrm{Im} \:%
\mathcal{G})$ severely limit the convergence of field-based simulations of
polymeric fluids.

In the present paper, we describe an alternative formulation of polymer
field theory that essentially side-steps the sign problem. Rather than
retaining just a Hubbard-Stratonovich chemical potential field for each
monomer species, we also retain a second density field. This two field
approach is not new \cite{matsen94,hongnoolandi81}, but rather is a usual
starting point for deriving SCFT. However, we show in the present paper that
at a fixed value of the density field, the functional integral over the
chemical potential field can be performed in the \emph{saddle point
approximation, and that this approximation is exact in the thermodynamic
limit of large systems.} This we believe to be a new result and one that has
profound consequences for field-theoretic computer simulations of polymeric
fluids.

\section{Basic Formalism: Polymer Solutions}

\subsection{Transformation to a Field Theory}

Consider a solution of $M$ homopolymer chains in a volume $V$, each chain
composed of $N$ monomers of size $a$. The solvent is not considered
explicitly, the temperature is denoted by $T$, and $\beta=1/T$. The
formalism presented below can be easily generalized to blends of different
chains and to melts or solutions of copolymers as will be seen later.

We assume that the monomers interact through a 2-body interaction potential
(of mean force) denoted by $v(r)$. In the celebrated Edwards model, the
interaction $v(r)$ is modelled as a Dirac delta function, $\delta (r)$, but
in general, the interaction $v$ can represent both solvent effects,
electrostatic effects or any other effective monomer-monomer interaction.
The canonical partition function of the system reads

\begin{equation}
\mathcal{Z}=\frac{1}{M!}\int\dprod\limits _{k=1}^{M}\mathcal{D}r_{k}\; \exp\left(-%
\frac{3}{2a^{2}}\sum_{k=1}^{M}\int_{0}^{N}ds\left(\frac{dr_{k}}{ds}%
\right)^{2}-\frac{\beta}{2}\sum_{k,l=1}^{M}\int_{0}^{N}ds\int_{0}^{N}ds^{%
\prime}v(r_{k}(s)-r_{l}(s^{\prime}))\right)  \label{part1}
\end{equation}
where $\int \mathcal{D}r_{k}$ denotes a path integral over all conformations
of the $k$th polymer chain.

Next, we introduce a monomer (number) density field $\rho (r)$ as a new
integration field. The microscopic expression for the monomer density, $\hat{%
\rho} (r)$, is given by
\begin{equation}
\hat{\rho}(r)=\sum_{k=1}^{M}\int_{0}^{N}ds~\delta(r-r_{k}(s))  \label{rhodef}
\end{equation}
If we constrain $\rho (r)$ to $\hat{\rho}(r)$ at each point $r$ of space by
the use of a $\delta$-function, the partition function of the polymer system
can be written as
\begin{equation}
\mathcal{Z}=\int\mathcal{D}\rho \;\exp\left(-\frac{\beta}{2}\int
drdr^{\prime}\rho(r)v(r-r^{\prime})\rho(r^{\prime})\right)Z(\rho)
\label{origin}
\end{equation}
where the function $Z(\rho)$ is defined by
\begin{equation}
Z(\rho) = \frac{1}{M!}\int\dprod\limits _{k=1}^{M}\mathcal{D}r_{k}
\exp\left(-\frac{3}{2a^{2}}\sum_{k=1}^{M}\int_{0}^{N}ds\left(\frac{dr_{k}}{ds%
}\right)^{2}\right) \dprod\limits _{r}\delta\left(\rho(r)-\hat{\rho} (r)
\right)  \label{zrho}
\end{equation}
and $\int \mathcal{D} \rho$ denotes a functional integral over the real,
scalar density field $\rho (r)$. The object $Z(\rho )$ is the partition
function of a set of $M$ continuous Gaussian chains of $N$ monomers,
constrained in such a way that the total monomer density is fixed to $%
\rho(r) $ at any point $r$ of space. This partition function evidently
counts the total number of chain conformations that are consistent with the
density profile $\rho(r)$. We thus write
\begin{equation}
Z(\rho)=\exp[S(\rho)]
\end{equation}
and identify $S(\rho)$ with the entropy of a system of $M$ chains
constrained to have a density profile $\rho(r)$. From its very definition,
it is clear that $Z(\rho)$ is real and positive for any real $\rho(r)$. In
addition, we expect $Z(\rho)$ to scale with $V$, $N$, and $M$ as $(V
g^{N})^M /M!$, where $V$ is the volume of the system and $g$ is some
positive, real constant.

At this stage, it is clear that we have expressed the partition function of
the polymeric system in terms of a functional integral over a real field
with a positive definite Boltzmann weight. We have yet to show how to
compute the entropic part of the weight, namely the factor $Z(\rho)$.

Using a Fourier representation of the delta function, the above expression
for $Z(\rho )$ can be written as
\begin{equation}
Z(\rho)=\frac{1}{M!}\int\mathcal{D}\phi\exp\left(i\int dr\phi(r)\rho(r)+M\ln
Q(i\phi)\right)  \label{z}
\end{equation}
where
\begin{equation}
Q(i\phi)=\int\mathcal{D}r\, e^{-\frac{3}{2a^{2}}\int_{0}^{N}ds\left(\frac{dr%
}{ds}\right)^{2}-i\int_{0}^{N}ds\phi(r(s))}  \label{zz}
\end{equation}
is the partition function of a \emph{single} polymer in the presence of the
potential $i \phi(r)$. It is important to note that this is a \emph{purely
imaginary potential} because the functional integral in eq.~(\ref{z}) is
over a real field $\phi (r)$. We recognize in eq.~(\ref{zz}) a Feynman path
integral for a quantum mechanical problem with quantum Hamiltonian $\mathcal{%
H}=-\frac{a^{2}}{6}\nabla^{2}+i\phi(r)$. Using standard quantum mechanical
notation, we may write
\begin{equation}
Q(i\phi)=\int dr\int dr^{\prime}<r|e^{-N\mathcal{H}}|r^{\prime}>
\label{feynman}
\end{equation}

\subsection{Expansion about the Saddle Point}

Of particular interest is an asymptotic expansion of $Z(\rho)$ for
large $M$ to examine the thermodynamic limit of $M \rightarrow
\infty , V \rightarrow \infty , \; M/V \; \text{finite}$. For this
purpose, we perform a saddle point expansion of the $\phi$
integral. The first step is to write an equation for the
mean-field (saddle point) $\phi_{0} (r)$ that produces the desired
$\rho (r)$
\begin{eqnarray}
\rho(r) & = & - M\frac{\delta \ln Q(i \phi_0 )}{\delta (i \phi_0 (r))}
\label{rho1} \\
& = & \frac{M}{Q(i\phi_0 )} \int_0^N ds \; \Psi (r,s)\Psi (r,N-s )
\label{rho2}
\end{eqnarray}
Here, $s$ is denotes a contour location along the chain and the
``propagator'' $\Psi (r,s)$ satisfies the diffusion equation
(analogous to the Schr\"{o}dinger equation)
\begin{equation}
(\frac{\partial}{\partial s}-\frac{a^{2}}{6}\nabla^{2}+i\phi_{0}(r))%
\Psi(r,s)=0  \label{dif1}
\end{equation}
subject to the initial condition $\Psi (r,0)=1$. The single chain partition
function is related to the field $\Psi$ by
\begin{equation}  \label{eq9x}
Q(i \phi ) = \int dr \; \Psi (r,N)
\end{equation}

>From the structure of the above equations, we can make a few general
statements. First, a necessary condition for eq.~(\ref{rho1}) to have a
solution $\phi_0 (r)$ is that $\rho(r)$ should be positive at any point $r$
and that its integral over space should be equal to $MN$. It is also clear
from eqs.~(\ref{rho1}) and (\ref{rho2}) that a self-consistent solution may
exist with a real $i\phi _{0}(r)$, i.e. purely imaginary $\phi_0 (r)$. From
the structure of these equations, we see that if $\phi _{0}(r)$ is a
mean-field (saddle point) solution, one can add to it any \emph{constant}
and obtain a new solution. To lift this degeneracy of the saddle-point, we
may specify the value of the field $\phi _{0}$ at a specific point in space,
or its spatially-averaged value. Having made such a choice, it can be proven
that such a real solution $i \phi_0 (r)$ not only exists, but is a \emph{%
unique} solution of the above equations.
%
%

To perform the asymptotic expansion about the saddle point $\phi_0$ at
higher orders, we shift the integration variable $\phi$ by writing
\begin{equation*}
\phi(r)=\phi_{0}(r)+\chi(r)
\end{equation*}
The partition function then reads
\begin{equation}
Z(\rho)=Z_{0}(\rho)\int\mathcal{D}\chi \; \exp\left(i\int dr
\;\chi(r)\rho(r)+M\ln\mathcal{Q}(i\chi)\right)  \label{eq14}
\end{equation}
where
\begin{equation}
\mathcal{Q}(i\chi)=\frac{\int\mathcal{D}r\, e^{-\frac{3}{2a^{2}}%
\int_{0}^{N}ds\left(\frac{dr}{ds}\right)^{2}-i\int_{0}^{N}ds
\:\phi_{0}(r(s))-i\int_{0}^{N}ds \: \chi(r(s))}}{\int\mathcal{D}r\, e^{-%
\frac{3}{2a^{2}}\int_{0}^{N}ds\left(\frac{dr}{ds}\right)^{2}-i\int_{0}^{N}ds
\:\phi_{0}(r(s))}}  \label{shift}
\end{equation}
and $Z_{0}$ denotes the mean-field value of the partition function
\begin{equation}
Z_{0}(\rho)=\frac{1}{M!} \exp\left(i\int dr\phi_{0}(r)\rho(r)+M\ln
Q(i\phi_{0})\right)  \label{mf}
\end{equation}

Next, we expand the exponent in the integrand of eq.~(\ref{eq14}) in powers
of $\chi$. For this purpose, it is convenient to reexpress eq.~(\ref{shift})
as
\begin{equation}
\mathcal{Q} (i \chi ) = <e^{-i\int_{0}^{N}ds \: \chi(r(s))}>_{0}  \label{exp}
\end{equation}
where the notation $< ... >_{0}$ describes the expectation value over the
conformations of a single polymer subjected to the mean-field potential $i
\phi_0$. Expanding in powers of $\chi$ and rescaling the field $\chi$ by a
factor of $\sqrt{M}$, we obtain
\begin{equation}
Z(\rho)=Z_{0}(\rho)\int\mathcal{D\mathrm{\chi}} \; \exp\left(-\frac{1}{2}%
\int drdr^{\prime}\;
\chi(r)G_{2}(r,r^{\prime})\chi(r^{\prime})+W(\chi)\right)  \label{act}
\end{equation}
where
\begin{equation}
W(\chi)=\sum_{p=3}^{\infty}\frac{(-i)^{p}}{p!M^{p/2-1}}\int dr_{1}\ldots
dr_{p} \; \chi(r_{1})\ldots\chi(r_{p})G_{p}(r_{1},\ldots,r_{p})  \label{w}
\end{equation}
and $G_{p}(r_{1},\ldots,r_{p})$ denotes the $p$-point \emph{connected}
(cumulant) correlation function for the density of a \emph{single chain},
defined by
\begin{equation}
G_{p}(r_{1},\ldots,r_{p})=<\bar{\rho}(r_{1})\ldots\bar{\rho}(r_{p})>_{c}
\label{green}
\end{equation}
where $\bar{\rho} (r) =\int_0^N ds \; \delta (r -r(s) )$ is the single chain
microscopic density. This expectation value is taken with respect to the
conformations of a single chain in the mean-field according to eq.~(\ref{exp}%
) with the usual definition of connected averages \cite{amit84}.
Note that there is no linear term in $\chi$ appearing in the exponent of
eq.~(\ref{act}) since the action is stationary with respect to $\phi_{0}$.

The partition function of eq.~(\ref{act}) can be calculated in a
perturbation expansion in powers of $1/M$. This is most conveniently done by
the use of Feynman diagrams in which propagator lines are associated with
factors of $S_{2}(r,r^{\prime})=G_{2}^{-1} (r,r^{\prime})$ and $p$th-order
vertices are assigned factors of $[(-i)^p /M^{p/2 -1} ] G_p (r_1 , ... , r_p
)$. The result is that
\begin{equation}
Z(\rho)=Z_{0}(\rho)\exp\left(-\frac{1}{2}\mathrm{Tr}\ln G_{2}(r,r^{\prime})+%
\mathrm{Sum\, of\, all\, Connected\, Diagrams}\right)  \label{free}
\end{equation}
where the connected diagrams have one or more vertices of degree $p \geq 3$.
Since the vertices in eq.~(\ref{green}) are connected correlation functions,
all terms in the exponent of eq.~(\ref{free}) are extensive, that is they
are proportional to the volume $V$ of the system. The general expansion of $%
Z(\rho)$ thus takes the form
\begin{equation}
Z(\rho)=\exp\left[-\beta V(f_{0}+\frac{f_{1}}{M}+\frac{f_{2}}{M^{2}}+...)%
\right]  \label{expansion}
\end{equation}
where the $f_j$ are contributions to the free energy density that are \emph{%
independent} of $M$ and $V$ in the thermodynamic limit. Since all vertices
carry powers of $1/M$, the sum of all connected graphs is negligible in the
thermodynamic limit, and eq.(\ref{free}) yields
\begin{equation}
Z(\rho)\simeq Z_{0}(\rho)\exp\left(-\frac{1}{2}\mathrm{Tr}\ln
G_{2}(r,r^{\prime})\right)  \label{result}
\end{equation}
in the limit $M\rightarrow\infty$. We note that $G_2 (r,r^\prime )$ is a
functional of $\rho$, because it represents the 2-point connected
correlation function of a chain subjected to the mean-field potential $i
\phi_0$ and $\phi_0$ is in turn related to $\rho$ by the saddle point
condition eqs.~(\ref{rho1})-(\ref{rho2}).

>From the form of eq.~(\ref{result}), it would appear that there is
a correction to the mean-field value $Z_0 (\rho )$ of the
restricted density partition function $Z (\rho )$ that is
associated with Gaussian fluctuations in $\chi (r)$ about the mean
chemical potential field $\phi_0 (r)$.
Interestingly, these corrections from Gaussian field fluctuations play \emph{%
no role} in the statistical properties of the field $\rho$ for $M
\rightarrow \infty$. This is demonstrated by two independent routes in the
next section.

\subsection{Significance of the Fluctuation Term}

To examine the significance of the fluctuation term $\frac{1}{2}\mathrm{Tr}%
\ln G_{2}(r,r^{\prime })$, we return to the original formulation of the
field theory prior to the asymptotic expansion in $M$. From eq.~(\ref{origin}%
) it is clear that we can write an effective, real field theory for the $%
\rho $ field as
\begin{equation}
\mathcal{Z}=\int \mathcal{D}\rho \;\exp [-H(\rho )]  \label{ghf1}
\end{equation}%
where the effective Hamiltonian $H(\rho )$ is given by
\begin{equation}
H(\rho )=\frac{\beta }{2}\int drdr^{\prime }\;\rho (r)v(r-r^{\prime })\rho
(r^{\prime })-S(\rho )  \label{ghf2}
\end{equation}%
where $S(\rho )=\ln Z(\rho )$ is given in eq.~(\ref{z}). One way to simulate
such a field theory is through a Langevin stochastic dynamics \cite{parisi88}
for $\rho $. Note that the density field $\rho (r)$ is constrained to be
positive at each point $r$. A simple way to enforce this condition in the
functional integral (\ref{ghf1}) is to add an external potential which is
infinite when $\rho (r)$ is negative and $0$ when $\rho (r)$ is positive. A
practical realization of such a potential can be obtained by adding an
exponential function to the Hamiltonian, and define

\begin{equation*}
H^{\prime }(\rho )=H(\rho )+\int dr~e^{-\lambda \rho (r)}
\end{equation*}%
where $\lambda $ is a constant, chosen large enough to ensure the positivity
constraint.

If we keep in mind that the total number of monomers is fixed, equal to $MN$%
, we may use the conserved dynamics (model B) \cite{halperin_hohenberg} and
write%
\begin{eqnarray}
\frac{\partial }{\partial t}\rho (r,t) &=&\Gamma \nabla ^{2}  \frac{%
\delta H^{\prime }}{\delta \rho (r,t)}+\eta (r,t)  \notag \\
&=&\Gamma \nabla ^{2}  \left( \beta \int dr^{\prime }v(r-r^{\prime
})\rho (r^{\prime },t)-<i\phi (r,t)>_{\rho }-\lambda e^{-\lambda
\rho (r)}\right) +\eta (r,t) \label{langevin}
\end{eqnarray}%
where we have used
\bigskip
\begin{equation*}
\frac{\delta S(\rho )}{\delta \rho (r,t)}=\frac{\delta \log Z(\rho )}{\delta
\rho (r,t)}=<i\phi (r,t)>_{\rho }
\end{equation*}%
and $\eta (r,t)$ is a Gaussian, white noise with vanishing first moment, $%
\langle \eta (r,t)\rangle =0$, and second moment satisfying the
usual fluctuation-dissipation theorem for a conserved noise,
\begin{equation}
<\eta (r,t)\eta (r^{\prime },t^{\prime })>=-2\Gamma \nabla^2
\delta (r-r^{\prime })\delta (t-t^{\prime }) \label{nose}
\end{equation}%
The Laplacian in front of the r.h.s. of (\ref{langevin}) ensures that the
space integral of $\rho(r,t)$, which is equal to the total number of
monomers $MN$, remains constant in time. 
In the above equation, the expectation value $<i\phi (r)>_{\rho }$ denotes
an equilibrium average of the field $i\phi (r)$ at constrained $\rho (r)$
using the statistical weight contained in eq.~(\ref{z}) for $Z(\rho )$.

With the use of eq.~(\ref{act}), it is easily seen that
\begin{equation*}
<i\phi (r)>_{\rho }=i\phi _{0}(r)+\frac{i}{\sqrt{M}}<\chi (r)>_{\rho }
\end{equation*}%
where
\begin{equation*}
<\chi (r)>_{\rho }=\frac{\int \mathcal{D}\chi \;\chi (r)\exp \left( -\frac{1%
}{2}\int drdr^{\prime }\chi (r)G_{2}(r,r^{\prime })\chi (r^{\prime })+W(\chi
)\right) }{\int \mathcal{D\mathrm{\chi }}\;\exp \left( -\frac{1}{2}\int
drdr^{\prime }\chi (r)G_{2}(r,r^{\prime })\chi (r^{\prime })+W(\chi )\right)
}
\end{equation*}%
and $W(\chi )$ is given by eq.~(\ref{w}). From this expression it follows
immediately that$_{{}}$
\begin{equation}
<i\phi (r)>_{\rho }=i\phi _{0}(r)+O(\frac{1}{M})  \label{onem}
\end{equation}%
and thus, in the Langevin equation formalism, it is sufficient to evaluate $%
<i\phi (r)>_{\rho }$ at the mean-field level to capture the leading
asymptotic behavior in the thermodynamic limit. The evaluation of the
Langevin forces on the right hand side of eq.~(\ref{langevin}) therefore
requires only the computation of $\phi _{0}(r,t)$, which follows from the
solution of the saddle point equations at prescribed $\rho (r,t)$. The
forces entering the Langevin equation are real (since $i\phi _{0}$ is real),
and thus we have a conventional real Langevin dynamics. The key result is
that in the thermodynamic limit, there is no signature of the fluctuation
term $(1/2)\mathrm{Tr}\ln G_{2}(r,r^{\prime })$ that appears in eq.~(\ref%
{result}) for $Z(\rho )$.

The same result can be directly obtained from eq.(\ref{result}) by means of
some straightforward manipulations. Indeed, from that equation and eq.~(\ref%
{mf}), we have
\begin{eqnarray}
\frac{\delta\ln Z(\rho)}{\delta\rho(r_{0})} & = & <i\phi(r_{0})>_{\rho} \\
& = & i\phi_{0}(r_{0})-\frac{1}{2}\frac{\delta~\mathrm{Tr~}\ln
G_{2}(r,r^{\prime})}{\delta\rho(r_{0})}
\end{eqnarray}
The $\mathrm{Tr} \ln$ term can be further simplified. Using some simple
identities, we get
\begin{eqnarray}
\frac{\delta~\mathrm{Tr~}\ln G_{2}(r,r^{\prime})}{\delta\rho(r_{0})} & = &
\int drdr^{\prime}\frac{\delta G_{2}(r,r^{\prime})}{\delta\rho(r_{0})}%
G_{2}^{-1}(r^{\prime},r)  \notag \\
& = & \int drdr^{\prime}dr_{1}\frac{\delta(i\phi_{0}(r_{1}))}{%
\delta\rho(r_{0})}\frac{\delta G_{2}(r,r^{\prime})}{\delta(i\phi_{0}(r_{1}))}%
G_{2}^{-1}(r^{\prime},r)  \label{eq30}
\end{eqnarray}
>From our previous definitions, we see that
\begin{equation}
\frac{\delta G_{2}(r,r^{\prime})}{\delta(i\phi_{0}(r_{1}))}
=-G_{3}(r,r^{\prime},r_{1})
\end{equation}
and the saddle point condition implies
\begin{equation}
\frac{\delta(i\phi_{0}(r_{1}))}{\delta\rho(r_{0})}=-\frac{1}{M}
G_{2}^{-1}(r_{1},r_{0})
\end{equation}
Substitution of these results into eq.~(\ref{eq30}) leads to a final
expression
\begin{equation}
\frac{\delta\ln Z(\rho)}{\delta\rho(r_{0})} = i\phi(r_{0})-\frac{1}{2M} \int
drdr^{\prime}dr_{1}G_{2}^{-1}(r_{1},r_{0})G_{3}(r,r^{
\prime},r_{1})G_{2}^{-1}(r^{\prime},r)  \label{correc}
\end{equation}
Again, since $G_{2}$ and $G_{3}$ are connected Green's functions, the triple
integral in eq.~(\ref{correc}) is finite in the thermodynamic limit (non
extensive) and this shows that to leading order in $1/M$, the expectation
value of $\phi$ is given by its mean-field value $\phi_{0}$, in agreement
with eq.~(\ref{onem}). In addition, it is easy to see that the correction
term calculated here is exactly the same as the one that would have been
obtained as the leading term of the perturbation expansion using eq.~(\ref%
{act}). We conclude that the fluctuation term $(1/2)\mathrm{Tr}\ln G_2
(r,r^\prime )$ can be safely omitted from the effective Hamiltonian $H(\rho
) $ when computing the Langevin force $\delta H/\delta \rho$ for the purpose
of simulating large systems. Moreover, it follows that the fluctuation
correction can also be dropped from the Hamiltonian itself, if Monte Carlo
simulations are to be employed.

We summarize with the statement that in the thermodynamic limit of large
systems, the field theory expressed in density variables can be written as
\begin{equation}
\mathcal{Z}=\int \mathcal{D}\rho \;\exp [-H^{\prime }(\rho )]
\label{eq34}
\end{equation}%
with
\begin{equation}
H^{\prime }(\rho )=\frac{\beta }{2}\int drdr^{\prime }\;\rho
(r)v(r-r^{\prime })\rho (r^{\prime })+\int dr~e^{-\lambda \rho (r)}-\int
dr\;i\phi _{0}(r)\rho (r)-M\ln Q(i\phi _{0})  \label{eq35}
\end{equation}%
and where the real field $i\phi _{0}(r)$ is obtained at a particular $\rho (r)$ by the solution of eq.~(%
\ref{rho1}).

\subsection{Langevin Dynamics and Monte Carlo: Implementation}

The Langevin equation (\ref{langevin}) can be discretized both in time and
space. A variety of time integration schemes for such nonlinear stochastic
differential equations are available \cite{kloeden}.

It is of interest to compare this \emph{real} Langevin scheme with the \emph{%
complex} Langevin scheme that has been used to numerically
simulate field theories of polymer solutions and melts expressed
in chemical potential [i.e. $\phi
(r) $] variables \cite{ganesan01,katz02}. Because the effective Hamiltonian $\mathcal{G%
}(\phi )$ in such field theories is complex, one must address the sign
problem. A useful strategy, originally devised by Parisi and Klauder \cite%
{parisi83,klauder84}, has been to extend the field variables to the complex
plane ($\phi=\phi_R + i \phi_I$) and write a Langevin dynamics
\begin{equation}  \label{eq37}
\frac{\partial}{\partial t} \phi_R (r,t) = - \Gamma \: \mathrm{Re} \frac{%
\delta \mathcal{G}}{\delta \phi (r,t)} + \eta (r,t)
\end{equation}
\begin{equation}  \label{eq38}
\frac{\partial}{\partial t} \phi_I (r,t) = - \Gamma \: \mathrm{Im} \frac{%
\delta \mathcal{G}}{\delta \phi (r,t)}
\end{equation}
where $\eta (r,t)$ is a real Gaussian, white noise with the
following covariance
\begin{equation}
<\eta (r,t)\eta (r^{\prime },t^{\prime })>=2\Gamma  \delta
(r-r^{\prime })\delta (t-t^{\prime })
\end{equation}
This procedure, if it converges \cite{gausterer98,lee94,schoenmaker87,ghf03}%
, will properly describe the stationary fluctuation spectrum of a model with
complex $\mathcal{G}(\phi )$. We have found it to be reasonably effective at
suppressing phase oscillations in simulations of polymer solutions and
melts, although sign problem is still very pronounced in dilute and
semidilute systems. An advantage of eqs.~(\ref{eq37})-(\ref{eq38}) is that
the computation of the complex force $\delta \mathcal{G}/\delta \phi $
requires only a \emph{single pass} at solving the diffusion equation (\ref%
{dif1}) (with $\phi _{0}\rightarrow \phi $), while computation of $\delta
\mathcal{H}^{\prime }/\delta \rho $ \ in the real Langevin scheme of eq.~(%
\ref{langevin}) requires multiple passes at solving the diffusion equation in
order to establish the field $\phi _{0}$ consistent with the current $\rho $%
. Whether the extra computational burden of the force evaluation
in the present density-based real Langevin scheme is offset by
avoidance of the sign problem remains to be seen and, indeed, may
prove to be system and problem dependent.

Note that in a Monte Carlo implementation, the minor complications
due to the positivity of $\rho (r)$ and the conservation of the
integral $\int dr\rho (r)$ are easy to overcome. First, one
discretizes space. To implement the conservation of the total
number of monomers, one starts with a configuration of $\rho (r)$
that has the right number of monomers, and then one chooses
randomly two lattice sites, say $r_{1}$ and $r_{2}$ . The
elementary Monte Carlo variation of $\rho (r)$ consists in
modifying simultaneously $\rho $ at two points $r_{1}$ and $r_{2}$
according to $\rho (r_{1})\rightarrow \rho (r_{1})+\delta \rho $
and $\rho (r_{2})\rightarrow \rho (r_{2})-\delta \rho $ . This
procedure obviously conserves the total integral of $\rho (r)$ in
space. Also, to enforce the positivity of $\rho (r) $, one just
has to reject any variation of $\rho $ that produces a negative
$\rho $.

\section{More complicated systems}

The concepts and methods presented above can be extended in a
straightforward way to virtually any type of inhomogeneous polymer system.
We illustrate in the context of a solution of AB diblock copolymers.

Let $f$ \ be the volume fraction of A monomers and $1-f$ the volume fraction
of B monomers on each diblock copolymer chain with a total of $N$ monomers.
Denoting by $v_{AA}(r)$, $v_{BB}(r)$, and $v_{AB}(r)$ the respective
potentials of mean force (mediated by the solvent) between the different
species of monomers, and using the same notations as in the previous
section, we may write
\begin{eqnarray}
\mathcal{Z} &=&\frac{1}{M!}\int \mathcal{D}\rho _{A}\mathcal{D}\rho _{B}\exp \left( -%
\frac{\beta }{2}\int drdr^{\prime }\rho _{A}(r)v_{AA}(r-r^{\prime })\rho
_{A}(r^{\prime })\right)   \notag \\
&\times &\exp \left( -\frac{\beta }{2}\int drdr^{\prime }\rho
_{B}(r)v_{BB}(r-r^{\prime })\rho _{B}(r^{\prime })\right)   \notag \\
&\times &\exp \left( -\beta \int drdr^{\prime }\rho
_{A}(r)v_{AB}(r-r^{\prime })\rho _{B}(r^{\prime })\right) Z_{A}(\rho
_{A})Z_{B}(\rho _{B})  \label{originco}
\end{eqnarray}%
1where the functions $Z_{A}(\rho _{A})$ and $Z_{B}(\rho _{B})$ are defined
by
\begin{eqnarray}
Z_{A}(\rho _{A}) &=&\int \dprod\limits_{k=1}^{M}\mathcal{D}r_{k}\exp \left( -%
\frac{3}{2a^{2}}\sum_{k=1}^{M}\int_{0}^{fN}ds\left( \frac{dr_{k}}{ds}\right)
^{2}\right)   \notag \\
&\times &\dprod\limits_{r}\delta \left( \rho
_{A}(r)-\sum_{k=1}^{M}\int_{0}^{fN}ds~\delta (r-r_{k}(s))\right)
\label{zrhoa}
\end{eqnarray}%
and
\begin{eqnarray}
Z_{B}(\rho _{B}) &=&\int \dprod\limits_{k=1}^{M}\mathcal{D}r_{k}\exp \left( -%
\frac{3}{2a^{2}}\sum_{k=1}^{M}\int_{fN}^{N}ds\left( \frac{dr_{k}}{ds}\right)
^{2}\right)   \notag \\
&\times &\dprod\limits_{r}\delta \left( \rho
_{B}(r)-\sum_{k=1}^{M}\int_{fN}^{N}ds~\delta (r-r_{k}(s))\right)
\label{zrhob}
\end{eqnarray}

It is clear from their definitions that both $Z_{A}$ and $Z_{B}$ are
positive definite, and using a Fourier representation for the delta
functionals, one can again show that in the limit of an infinite number of
chains (with infinite volume and finite concentration), the corresponding
``chemical potential'' field integrals can be asymptotically evaluated by
the saddle point method. The resulting expression for the entropy functional
\begin{equation}  \label{eq43}
S(\rho_A , \rho_B ) = \ln [Z_A (\rho_A ) Z_B (\rho_B )]
\end{equation}
is
\begin{equation}  \label{eq44}
S(\rho_A , \rho_B ) =\int dr \; ( i \phi_A \rho_A + i \phi_B \rho_B )+ M \ln
Q(i \phi_A , i \phi_B )
\end{equation}
where $Q(i \phi_A , i \phi_B )=\int dr \; \Psi (r,N)$. The propagator $\Psi
(r,s)$ satisfies (again, subject to $\Psi (r,0)=1$)
\begin{equation}  \label{eq45}
(\frac{\partial}{\partial s}-\frac{a^{2}}{6}\nabla^{2}+w(r,s))\Psi(r,s)=0
\end{equation}
where $w(r,s)=i \phi_A (r)$ for $s \in (0,Nf)$ and $w(r,s)=i \phi_B (r)$ for
$s \in (Nf,N)$. The density fields are related to the saddle point chemical
potentials, $\phi_A$ and $\phi_B$, by equations analogous to eq.~(\ref{rho2}%
)
\begin{equation}  \label{eq46}
\rho_A (r) = \frac{M}{Q(i\phi_A ,i\phi_B )} \int_0^{Nf} ds \; \Psi
(r,s)\Psi^* (r,N-s )
\end{equation}
\begin{equation}  \label{eq47}
\rho_B (r) = \frac{M}{Q(i\phi_A ,i\phi_B )} \int_{Nf}^{N} ds \; \Psi
(r,s)\Psi^* (r,N-s )
\end{equation}
A second propagator $\Psi^* (r,s)$ enters these expressions due to the
head-to-tail asymmetry of diblock copolymers. It satisfies
\begin{equation}  \label{eq48}
(\frac{\partial}{\partial s}-\frac{a^{2}}{6}\nabla^{2}+w^* (r,s))\Psi^*
(r,s)=0
\end{equation}
subject to $\Psi^* (r,0)=1$, where $w^* (r,s)=i \phi_B (r)$ for $s \in
(0,N(1-f))$ and $w^* (r,s)=i \phi_A (r)$ for $s \in (N(1-f),N)$.

A formalism very similar to this was employed recently, without
formal justification, for a real Langevin dynamics study of
spinodal decomposition in binary polymer blends \cite{reister} and
in studying microemulsion formation in ternary copolymer alloys
\cite{dominik}. The present analysis shows that this is indeed a
rigorous approach to simulating fluctuating systems of macroscopic
size.

\section{Summary and Discussion}

In the present paper we have examined statistical field theories of
polymeric fluids with the objective of identifying field-based models that
are well-suited to numerical simulation. It has been known for some time
that essentially any microscopic (particle-based) model of a polymeric fluid
can be converted to a field theory in density and chemical potential
variables \cite{matsen94,hongnoolandi81}, e.g. for a one-component system
\begin{equation}
\mathcal{Z}=\int \mathcal{D}\rho \int \mathcal{D}\phi \;\exp
[-\mathcal{H}(\rho ,\phi )]  \label{eq49}
\end{equation}%
where the effective Hamiltonian $\mathcal{H}(\rho ,\phi )$ is a \emph{complex%
} functional of the two fields. In the case of pair interactions
among monomers, $\mathcal{H}(\rho ,\phi )$ is a quadratic form in
$\rho $, so the density field can be integrated out exactly (for
potentials with an inverse) to obtain a statistical field theory
in $\phi $
\begin{equation}
\mathcal{Z}=\int \mathcal{D}\phi \;\exp [-\mathcal{G}(\phi )]
\label{eq50}
\end{equation}%
In this case, the effective Hamiltonian $\mathcal{G}(\phi )$ is
 complex. This type of field theory has been studied using
approximate analytical methods for many years \cite{edwards65} and
has recently been examined with complex Langevin simulation
techniques \cite{ganesan01,ghfrev,katz02}.
Unfortunately, the non-positive definite character of the Boltzmann factor $%
\exp (-\mathcal{G})$ gives the integrand an oscillatory character
(\textquotedblleft sign problem\textquotedblright ), which can dramatically
hinder the convergence of numerical simulations.

Rather than transforming to a field theory in the single field $\phi$, a
more typical strategy for studying inhomogeneous polymers is to invoke a
mean-field approximation. This amounts to evaluating both integrals in eq.~(%
\ref{eq49}) at leading order by the saddle point method. This reproduces the
familiar self-consistent field theory (SCFT) \cite{helfand75} and amounts to
\begin{equation}  \label{eq51}
\mathcal{Z} \approx \exp [-\mathcal{H}(\rho^*, \phi^* )]
\end{equation}
where $\rho^*$ and $\phi^*$ are the ``self-consistent'' density and chemical
potential fields that satisfy $\delta \mathcal{H}/\delta \rho^* =0$, $\delta
\mathcal{H}/\delta \phi^* =0$. This is a powerful approach, but it neglects
all field fluctuations, so the utility of the SCFT is limited to highly
concentrated systems of high molecular weight polymers, far from critical
points and phase transitions where soft fluctuation modes occur.

In the present paper, we have discussed a third approach in which the
chemical potential field $\phi$ is integrated out of eq.~(\ref{eq49}),
leading to a field theory solely in the monomer density
\begin{equation}  \label{eq52}
\mathcal{Z} = \int \mathcal{D} \rho \; \exp [-H(\rho )]
\end{equation}
This approach has the benefit of producing a \emph{real} field theory in
which $\exp (-H)$ is positive definite so that the sign problem is avoided
and standard Monte Carlo and real Langevin simulation methods can be
applied. In the past it was assumed that the $\phi$ integral necessary to
reduce eq.~(\ref{eq49}) to eq.~(\ref{eq52}) was intractable, leading various
authors to impose a variety of approximations in order to construct the
density functional $H(\rho )$ \cite{leibler80,tang91}. We have demonstrated
here, however, that the $\phi$ integral in eq.~(\ref{eq49}) can be \emph{%
exactly} evaluated by the saddle point method provided that the
thermodynamic limit is taken. The resulting real energy functional
$H(\rho )$ is given (for a homopolymer solution) by
eq.~(\ref{eq35}) and involves the constrained (real) saddle point
$i \phi_0 (r)$, which is the mean chemical potential field
required to generate a particular density pattern $\rho (r)$.

We believe that our discovery will have a number of significant
implications. First of all, it provides theoretical justification for
\textit{ad hoc} approaches that have been used to include thermal
fluctuations in density field-based simulations of polymers \cite%
{reister,fraaije96}. Secondly, by circumventing the sign problem, it avoids
the difficulties that have been encountered in field-theoretic polymer
simulations based on eq.~(\ref{eq50}). Moreover, with a real field theory, a
much wider variety of options exist for carrying out stochastic computer
simulations. Finally, we note that our proof that the $\phi$ integral in
eq.~(\ref{eq49}) is dominated for large systems by the saddle point carries
over to a much broader class of classical fluids, both simple and complex.
For example, real density functionals can be constructed using the present
methods for alternative polymer chain models, such as the worm-like chain
used to describe semiflexible polymers. Real density-based field theories
for a wide variety of polymers, copolymers, and alloys can be derived,
including solutions of charged polyelectrolyte systems. In the case of
simple fluids, we believe that the present formalism will provide new
insights and numerical strategies for improving density functional methods,
including the treatment of crystallization and melting.

A clear disadvantage of the density field approach, at least for
polymers, is that force or energy evaluations require the
computation of the mean-field $i \phi_0 (r)$ from equations such
as (\ref{rho2}) at constrained monomer density $\rho (r)$. This
must be done iteratively and each iteration requires an
independent solution of the diffusion equation (\ref{dif1}). In
constrast, evaluation of the energy $\mathcal{G} (\phi )$ in the
potential field formulation of eq.~(\ref{eq50}) requires only a
single solution of the diffusion equation. Since the vast majority
of the computational effort in a field-theoretic computer
simulation of a polymeric fluid is spent in solving the diffusion
equation, it is essential that a very efficient scheme be devised
for computing $i \phi_0$ given $\rho$. We are cautiously
optimistic, however, that the extra computational burden
associated with simulating in the density variables will be more
than offset by the avoidance of the sign problem.

\begin{acknowledgments}
This work was supported by the National Science Foundation under
award DMR 03-12097. One of the authors (H.O.) wishes to thank the
MRL at UCSB for its kind hospitality during the course of this
work.
\end{acknowledgments}

\newpage

\bibliographystyle{unsrt}
\bibliography{paper}

\end{document}